# The quadrupole mechanism of hole spin relaxation.

Yuri A. Serebrennikov

Qubit Technology Center

2152 Merokee Dr., Merrick, NY 11566

ys455@columbia.edu

For spin-3/2 holes the anisotropic part of the instantaneous Luttinger Hamiltonian can be represented as an effective quadrupole coupling. We investigate the hole spin relaxation process induced by nonadiabatic fluctuations of this interaction. The obtained analytical solution of the stochastic Liouville equation describes the polarization decay of spin-3/2 holes in all regimes of momentum scattering: from collision-dominated to ballistic. Our results create the basis for quantitative interpretation of recent experiments and elucidate the striking difference between the hole spin relaxation process in bulk crystals and 2D semiconductor nanostructures.

72.25.Rb  71.55.Eq

## I. Introduction

The discovery of hole-mediated ferromagnetism[1] and ultra-fast demagnetization (< 1 ps) of III-Mn-V materials[2], drastic increase in the heavy hole spin lifetime (~ 270 $\mu s$) in quantum dots[3] (QDs), and observation of spin-Hall effect in *p*-type semiconductors[4] attracted considerable attention to hole spin dynamic and relaxation[5]. The hole spin does not couple through Fermi contact interaction to lattice nuclei and, hence, is free from the hyperfine channel of decoherence, which is very efficient for electron spin at zero and low magnetic fields[6]. On the other hand, due to the *p*-symmetry of the valence band, the intrinsic spin-orbit coupling (SOC) is much stronger for holes than for electrons. This leads to a strong mixing between the hole spin and orbital degrees of freedom. Recently it has been shown that, even in spherical bands and zero magnetic fields, the resultant total angular momentum $\vec{J}$ of a hole may exhibit a fast precession due to the action of an effective *quadrupole coupling*, which represents the anisotropic part of instantaneous Luttinger Hamiltonian[7]. In bulk crystals, the effective quadrupole Hamiltonian determines the splitting, $\Delta_{HL}$, between heavy hole (HH) and light hole (LH) subbands and depends on the wandering of a hole lattice momentum $\vec{k}$ in the angular space[8]. Thermal motion of a hole results in stochastic modulation of this interaction and may induce *nonadiabatic* transitions between the HH and LH subbands, which lead to *J*-relaxation and dephasing[7,8,9].

Clearly, this mechanism of hole spin relaxation is qualitatively similar to Dyakonov-Perel (DP) mechanism of electron spin relaxation. Yet the spin relaxation of holes in bulk crystals was discussed mostly in the context of the Elliot-Yafet (EY) spin dephasing process[10,11], where the phase modulation of heavy and light holes happens during *adiabatic* momentum scattering events. For holes, the DP mechanism has been considered only in 2D systems[12]. Recently, the hole spin dephasing due to stochastic modulation of the effective magnetic field (Rashba SOC term) in 2D quantum wells with large separation between LH and HH subbands was studied by constructing and numerically solving the kinetic spin Bloch equations[13]. The study goes beyond the usual Born-Redfield approximation and takes into account the effects of the inhomogeneous broadening and Coulomb scattering. It has been shown that for hole densities $>10^{11} cm^{-2}$ the carrier-carrier scattering may notably affect the hole spin dephasing process.

Since $\Delta_{HL}$ is orders of magnitude stronger than the effective magnetic fields due to broken inversion symmetry, the DP-like quadrupole mechanism can yield the main contribution to the hole spin-3/2 relaxation and lead to the ultra-fast relaxation in bulk crystals. The rapid progress of polarization- and time-resolved femtosecond spectroscopy allows direct measurements of these extremely short, on the order of 0.1 ps, relaxation times[14]. Theoretically, hole spin relaxation in the sub-picosecond time-range was studied within the framework of non-Markovian stochastic theory[8], which explicates the containment of nonadiabatic transitions between $\Gamma_7$ and $\Gamma_8$ multiplets by strong intrinsic SOC in III-V semiconductors. Adiabatic modulation of the $\Gamma_7$-$\Gamma_8$ splitting by elastic scattering of $\vec{k}$ does not lead to dephasing of the split-off holes. Nevertheless, the same perturbation can be essentially nonadiabatic relative to the splitting between the HH and LH bands inside the $\Gamma_8$ quadruplet. It has been shown that in the collision-dominated regime, $\Delta_{HL}^2 \tau_2^2 \ll 1$, where $\tau_2$ is the orientational relaxation time of the second rank quadrupole tensor, one cannot distinguish between the HH and the LH subbands. In this limit, the angular displacement of $\vec{J}$ turns into the process of small random walks in the angular space and the kinetics of $J$-relaxation is purely exponential (no oscillations). If the band structure and $\vec{k}$-scattering are isotropic, the rate of this process is determined by[7][8][9]

$$1/\tau_J(\Gamma_8) = (2/5)\Delta_{HL}^2 \tau_2 \ . \qquad (1)$$

Notably, Abraham and Pound[15] considered the formally equivalent problem more than fifty years ago in the context of nuclear spin-3/2 relaxation at zero magnetic fields and, with the obvious redefinition of the parameters, arrived to the same result. It follows from Eq.(1) that, similar to DP mechanism of electron spin relaxation, $\tau_J$ is inversely proportional to $\tau_2$. This prediction, however, requires unrealistically short $\tau_2$ to wash out the LH-HH splitting of about 40 meV, which represents $\Delta_{HL}$ in GaAs at room temperatures[16]. In fact, the distinct optical orientation and relaxation of HH's and LH's observed by Hilton and Tang[14] in undoped bulk GaAs clearly demonstrates that the system under study was outside the collision-dominated regime.

In general, if a *k*-dependent interaction that determines the spin-dynamics of a charge carrier at zero magnetic fields is larger than the momentum scattering rate, the Born-Redfield approximation (Abragam-Pound method) is not valid and the conventional theory of the DP mechanism of spin relaxation is not applicable. The kinetic theory can be extended to the range of strong interactions, $\Delta_{HL}^2 \tau_2^2 > 1$, only under certain assumptions regarding the random process. Here we assume that the change in the orientation of the principal axes of an effective quadrupole tensor can be modeled by the purely discontinuous Markovian process. Physically this means that the random direction of these axes varies instantaneously at the moments of successive collisions and is constant within the time intervals between them. Apparently, the validity of this model requires nonadiabaticity of a scattering event, which is the good approximation for systems with $\Delta_{HL}$ < 100 mEv. This approach allows us to employ the formalism of sudden modulation theory and stochastic Liouville equation[17][18] to calculate the response of $\vec{J}$ to instantaneous modulation of the quadrupole Hamiltonian even outside the collision-dominated regime. The obtained analytical solution describes the dumping of $\vec{J}$ precession in all regimes of momentum scattering: from fast ($\Delta_{HL}^2 \tau_2^2 \ll 1$) to slow modulation ($\Delta^2_{HL}\tau^2_2 > 1$) and ballistic ($\Delta^2_{HL}\tau^2_2 \to \infty$). Within the collision-dominated regime, our general expression yields the result of the stochastic perturbation theory, Eq.(1). In the opposite limit, our results describe a rather complex relaxation kinetics that reflects dumped oscillations in the population of heavy and light holes. Finally, our study demonstrates that the drastic drop in the rate of the hole spin relaxation in low-dimensional semiconductor nanostructures in comparison to bulk crystals can be explained by the 2D-confinement of the hole motion, which leads to suppression of the quadrupole mechanism.

### II. Theory

The spin-dynamics of $J = 3/2$ holes is determined by the 4x4 matrix of the instantaneous Luttinger Hamiltonian[19]. It has been shown[8] that within the "spherical approximation"[20][21] this Hamiltonian can be written in the following form (columns below correspond to m = 3/2, ½, -1/2, -3/2)

$$<3/2m_1;\vec{k}^{(M)}|H^{(M)}_{k^2}|3/2m;\vec{k}^{(M)}> = \frac{\gamma_1 k^2}{2m_0}\delta_{mm_1} + \frac{\gamma_2}{m_0}\begin{pmatrix} D_k & 0 & \sqrt{3}E_k & 0 \\ 0 & -D_k & 0 & \sqrt{3}E_k \\ \sqrt{3}E_k & 0 & -D_k & 0 \\ 0 & \sqrt{3}E_k & 0 & D_k \end{pmatrix}.$$

Here the superscript (*M*) denotes the principal-axes system of the effective quadrupole tensor $\vec{Q}_{ij} = [L_i L_j + L_j L_i - (2L^2/3)\delta_{ij}]/2$, where $i,j = x_L, y_L, z_L$ represent the Cartesian basis in the space-fixed lab (*L*) frame, $\vec{L}$ is the effective orbital angular momentum of a hole, $\gamma_1$ and $\gamma_2$ are the dimensionless Luttinger parameters, $m_0$ is the bare electron mass, $D_k := -(2k^2_{z_M} - k^2_{x_M} - k^2_{y_M})/2$ and $E_k := -(k^2_{x_M} - k^2_{y_M})/2$. For $J = 3/2$ this matrix can be represented in terms of the irreducible spin-tensor operators of the full rotation group $T_{2q}(J) = (5/4)^{1/2} \sum_{m m_1} C^{Jm_1}_{Jm 2q} |Jm_1\rangle\langle Jm|$ as

$$H^{(M)}_{k^2}(J) = H_0 + V^{(M)}_Q = (\gamma_1/2m_0)k^2 + (6)^{1/2}(\gamma_2/m_0)\sum_q (-1)^q K^{(M)}_{2q} T^{(M)}_{2-q}(J), \quad (2)$$

where $C^{2q}_{1\mu 1\mu_1}$ denotes the Clebsch-Gordon coefficient[22], $K^{(M)}_{20} = (2/3)^{1/2} D_k$, $K^{(M)}_{2\pm 1} = 0$, and $K^{(M)}_{2\pm 2} = E_k$. Here $H_0$ denotes the isotropic part of the kinetic energy of a hole, $V^{(M)}_Q$ stands for the effective quadrupole interaction, which represents the coupling between $\vec{J}$ and the lattice momentum and is clearly anisotropic, $[J^2, V^{(M)}_Q] \neq 0$. Physically this means that that the translational motion of a hole *locally* breaks the isotropy of the system and, similar to a crystal field, lifts the degeneracy of the $\Gamma_8$ ($|3/2 m; k = 0\rangle$) "fine-structure" states that may exist only at the $\Gamma$-point. In the axially symmetric case ($E_k = 0$) the matrix of $H^{(M)}_{k^2}$ is diagonal in the $|3/2 m; \vec{k}\rangle$ basis, $J_{z_M}$ is conserved, and the eigenfunctions of $H^{(M)}_{k^2}$ can be classified by the helicity $m = \hat{k} \cdot \vec{J}$. Bands with $m = \pm 3/2$ correspond to HHs, while bands with $m = \pm 1/2$ represent LHs with $\Delta_{HL} = 2\gamma_2 D_k / m_0$. Due to the $T$-invariance of $H^{(M)}_{k^2}$, each of these subbands has Kramers' degeneracy. Note that for holes moving along $z_M$, Eq.(2) can be written in the familiar form $H^{(M)}_{k^2} = (k^2/2m_0)[\gamma_1 + 2\gamma_2(J^2/3 - J^2_{Z_M})]$. Thus, even if the carrier equilibrium distribution in the *k*-space is isotropic, the *instantaneous* Luttinger Hamiltonian outside the zone center lacks spherical symmetry. Random scatterings of $\vec{k}$ results in the random modulation of $V_Q$, which connects the tightly coupled *L-S* subsystem (a hole) to the lattice and is, therefore, responsible for inter-subband transitions and $\vec{J}$-relaxation. The main advantage of the expansion (2) is simplicity of the transformation of irreducible tensor operators $T_{2q}(J)$ under rotations of the coordinate system[22],

$$H_{k^2}^{(L)}(\Omega_t) = D^{-1}(\Omega_t) H_{k^2}^{(M)} D(\Omega_t) = \frac{\gamma_1}{2m_0} k^2 + \frac{\sqrt{6}\gamma_2}{m_0} \sum_{qp} (-1)^p T_{2p}^{(L)}(J) D^2_{q,-p}(\Omega_t) K_{2q}^{(M)}, \quad (3)$$

which significantly simplifies the theoretical study of the $\vec{J}$-relaxation presented below. Here $D(\Omega_t)$ is the operator of finite rotation, $\Omega_t = \{\alpha_t, \beta_t, \gamma_t\}$ is the set of Euler angles that represents the instantaneous orientation of the *L*-frame relative to the *M*-frame of reference at the moment *t*, $D^2_{q,-p}(\Omega_t)$ is the corresponding Wigner rotation matrix.

The basic problem is the calculation of the response of $\vec{J}$ to a random realization of

$$V_Q^{(L)}(\Omega_t) = \frac{\sqrt{6}\gamma_2}{m_0} \sum_{qp} (-1)^p T_{2p}^{(L)}(J) D^2_{q,-p}(\Omega_t) K_{2q}^{(M)}. \quad (4)$$

Without a precise definition of this process, this goal can be achieved only in the collision-dominated regime, where the stochastic perturbation is smaller than the inverse of the relevant correlation time. Henceforth, we shall assume that the main source of the stochastic time dependence of the effective quadrupole interaction Eq.(4) is the purely discontinues Markovian process in which the principal axes of the quadrupole tensor have a fixed direction for a mean time $\tau_Q$ and then jump instantaneously to a new orientation at successive times over a Poisson distribution. It is evident that such a process is a model. The time interval in which the variation of the Hamiltonian takes place ("collision" time) $\tau_c$ must be finite (~10 fs) even if short compared to $\tau_Q$. We can neglect the shortest of the times if this change is nonadiabatic $\|V_Q\|\tau_c \ll 1$ and, hence, the intricate details of a collision are unimportant. In this case, the orientation of $\vec{J}$ is the same immediately after the jump of the *M*-frame in the angular space. As a result, even if $\vec{J}$ is parallel to $z_M$ during some interval $t_{i+1} - t_i$, it will not commute with $V_Q$ after a sudden change of $\Omega$ and begin to precess about the new direction of an effective quadrupole field. It then follows that in the Heisenberg representation the partially averaged operator $\vec{J}^{(L)}(\Omega,t)$ obeys the stochastic Liouville equation of motion[17][18] ($\hbar = 1$):

$$\dot{\vec{J}}^{(L)}(\Omega,t) = i[V_Q^{(L)}(\Omega), \vec{J}^{(L)}(\Omega,t)] - \tau_Q^{-1}[\vec{J}^{(L)}(\Omega,t) - \int f(\Omega,\Omega')\vec{J}^{(L)}(\Omega',t)d\Omega']. \quad (5)$$

Here $\Omega'$ and $\Omega$ describe the orientation of the principal axes of the effective quadrupole tensor before and after the jump, correspondingly. The degree of correlation at the energy $\varepsilon_k$ is determined by the function $f(\Omega,\Omega')$. The latter depends only on the angle between the successive directions of the *M*-frame, is normalized, and conserves the stationary angular distribution $\varphi(\Omega) = \int f(\Omega,\Omega')\varphi(\Omega')d\Omega'$. If $f(\Omega,\Omega')$ is close to $\delta$-function, the random values change negligibly at every jump, the process is strongly correlated ("weak collision" or orientational diffusion limit). If $\varphi(\Omega)$ is re-established after every jump, $f(\Omega,\Omega') = \varphi(\Omega)$, the process is uncorrelated ("strong collision" limit). Equation (5) must be solved with the initial condition

$$\vec{J}^{(L)}(\Omega,0) = \varphi(\Omega)\vec{J}^{(L)}(0) \quad (6)$$

and the final physical information can be extracted from the ordinary integral of the solution over the angular space. Consequently, the autocorrelation function of $J_{Z_L}$ is given by

$$K_{J_{Z_L}}(t) := Tr \rho_{eq}^{(L)} \int J_{Z_L}(\Omega,t) J_{Z_L}^+(0) d\Omega, \qquad (7)$$

where $\rho_{eq}^{(L)}$ is the equilibrium density operator. It is important to note here that due to the assumed isotropy of either bulk crystals or 2D nanostructures (in-plane isotropy), all directions of an effective quadrupole tensor are equiprobable. Therefore, the conditional probability density $f(\Omega,\Omega')$ depends only on the angle $\tilde{\Omega} = \Omega - \Omega'$ between the successive directions of the *M*-frame, i.e.,

$$f(\Omega,\Omega') = f(\tilde{\Omega}). \qquad (8)$$

This is a standard formulation of the problem in the sudden modulation theory. Equation (5) is rather general and provides the computational bridge between the spin relaxation of a charge carrier and random wandering of its crystal momentum in the angular space. It is mathematically closed, however, rather complex. Evidently, for spin-3/2 holes (electron spin relaxation will be considered elsewhere) Eqs.(4) - (7) are formally equivalent to the problem of nuclear spin-3/2 relaxation at zero magnetic fields[23]. Remarkably, the property of the kernel (8) is sufficient to advance in solving Eq.(5). It has been shown[24] that Eq.(5) can be reduced to a differential one, which is formally identical to the master equation of the impact theory (see Appendix):

$$\dot{\tilde{J}}_q^{(M)}(t) = i \hat{L}_{k^2}^{(M)}(0) \tilde{J}_q^{(M)}(t) - \tau_Q^{-1} [\tilde{J}_q^{(M)}(t) - \sum_{q_1} \hat{T}_{q q_1} \tilde{J}_{q_1}^{(M)}(t)]. \qquad (9)$$

Here we introduce the following designations

$$\tilde{J}_q^{(M)}(t) = \int D(\Omega) J_q^{(L)}(\Omega,t) D(-\Omega) D_{q0}^1(\Omega) d\Omega, \tag{10}$$

$$J_{q=0} = J_{Z_L}, \quad J_{q=\pm 1} = \mp (J_{X_L} \pm i J_{Y_L})/\sqrt{2}. \tag{11}$$

The Liouvillian

$$\hat{L}_{k^2}^{(M)}(0) \tilde{J}_q^{(M)}(t) = [V_Q^{(M)}(0), \tilde{J}_q^{(M)}(t)] \tag{12}$$

determines the dynamic evolution of the system in the *M*-frame, while the action of the collision operator

$$\hat{T}_{qq_1} \tilde{J}_{q_1}^{(M)}(t) = \int f(\tilde{\Omega}) D(\tilde{\Omega}) \tilde{J}_{q_1}^{(M)}(t) D(-\tilde{\Omega}) D_{qq_1}^1(\tilde{\Omega}) d\tilde{\Omega} \tag{13}$$

is reduced to a linear transformation of the $J_q$-components in the Liouville-space.

The autocorrelation function Eq.(7) can be expressed via the solution of Eq.(9)

$$K_{J_{Z_L}}(t) = Tr \sum_q \rho_{eq}^{(M)} \tilde{J}_q^{(M)}(t) J_q^{(L)}, \tag{14}$$

which must be solved with the initial condition

$$\tilde{J}_q^{(M)}(t) = J_q^{(L)^+}/3. \tag{15}$$

After straightforward algebra it can be shown that calculation of the spectral function

$\tilde{K}_{J_Q}(\omega) = \pi^{-1} \text{Re} \int_0^\infty K_{J_Q}(t) \exp(-i\omega t) dt$ requires the inversion of the *finite* matrix:

$$\tilde{K}_{J_Q}(\omega)/K_{J_Q}(0) = (\pi)^{-1} \text{Re} < 1,0,0 | \frac{1}{-i(\omega \hat{1} + \hat{\Lambda}) + \hat{\Gamma}} | 1,0,0 >, \tag{16}$$

where the elements of the evolution $\hat{\Lambda}$ and the relaxation $\hat{\Gamma}$ operators are determined as follows

$$< K,l,q | \hat{\Lambda} | K',l',q' > = \Pi_{KK'll'} \chi(lq,l'q')[(-1)^{K+K'} - 1] \begin{Bmatrix} 3/2 & 3/2 & 3/2 \\ K & K' & 2 \end{Bmatrix} \begin{Bmatrix} l' & l & 2 \\ K & K' & 1 \end{Bmatrix}, \tag{17}$$

$$< K,l,q | \hat{\Gamma} | K',l',q' > = W_{qq'}^l \delta_{KK'} \delta_{ll'}. \tag{18}$$

Here $\Pi_K := (2K+1)^{1/2}$,

$$W_{qq'}^l = \tau_Q^{-1}[\delta_{qq'} - \int f(\widetilde{\Omega})D_{qq'}^l(\widetilde{\Omega})d\widetilde{\Omega}], \quad (19)$$

$$\chi(lq,l'q') = (-1)^{q'}[\sqrt{6}D_k C_{l'q' lq}^{20} + E_k(C_{l'q' lq}^{22} + C_{l'q' lq}^{2-2})]. \quad (20)$$

It is easy to see from Eqs.(17) - (20) that if the effective quadrupole interaction is axially symmetric, the problem reduced to inversion of the 5x5 matrix, which yields the following analytical expression for the normalized spectral function[24]

$$\widetilde{K}_{J_0}(\omega)/K_{J_0}(0) = (1/\pi)\operatorname{Re}\frac{A(\omega)}{i\omega A(\omega) + B(\omega)}, \quad (21)$$

$$A(\omega) = \frac{3}{7}\Delta_{HL}^2(i\omega + \tau_4^{-1}) + \frac{6}{35}\Delta_{HL}^2(i\omega + \tau_2^{-1}) + (i\omega + \tau_2^{-1})^2(i\omega + \tau_4^{-1}), \quad (22)$$

$$B(\omega) = \frac{2}{5}\Delta_{HL}^2(i\omega + \tau_2^{-1})(i\omega + \tau_4^{-1}). \quad (23)$$

Here $1/\tau_l = (1/\tau_Q)\int_{-1}^{1} f(\cos\widetilde{\beta})[1 - P_l(\cos\widetilde{\beta})]d(\cos\widetilde{\beta})$ is the inverse orientational relaxation time of the *l*-rank tensor and $P_l$ is the Legendre polynomial. It follows from Eqs.(21) - (23) that in general the spectral function depends on both times $\tau_2$ and $\tau_4$, which are rather different in the "weak collision" limit: $\tau_l^{-1} = l(l+1)D_\beta$, where $D_\beta$ is the coefficient of orientational diffusion ($\tau_4 = 0.3\tau_2$). In contrast, in the "strong collision" limit $f(\cos\widetilde{\beta}) = 1/2$ and $\tau_4 = \tau_2 = \tau_Q$. In the slow modulation regime, $\Delta_{HL}^2\tau_2^2 \gg 1$, Eqs.(21)-(23) describe the well resolved triplet structure in the spectral function, which reflects fast coherent HH-LH inter-subband oscillations (precession of $\vec{J}$ in the quadrupolar field) $K_{J_{Z_L}}(t)/K_{J_{Z_L}}(0) = [3 + 2\cos(\Delta_{HL}t)]/5$, dumped by the relaxation process. The integral rate of this process, defined as $\tau_J(\Gamma_8) = \pi \widetilde{K}_{J_0}(\omega \to 0)/K_{J_0}(0)$, is given by

$$\tau_J^{-1}(\Gamma_8) = \frac{14}{15}\tau_2^{-1}[1+\frac{2\tau_4}{5\tau_2}]^{-1} = \begin{cases} \approx 6D_\beta, & \text{"weak collisions"} \\ \tau_Q^{-1}, & \text{"strong collisions"} \end{cases}, \quad (24)$$

which does not depend on Luttinger parameters and has a negligible dependence on $\tau_4$ even in the "weak collision" limit. In the collision-dominated regime, the spectral function Eq.(21) has a simple Lorentzian form that corresponds to the pure exponential decay of hole spin polarization with the rate

$$\tau_J^{-1}(\Gamma_8) = \frac{8}{5}(\frac{\gamma_2}{m_0}D_k)^2 \tau_2 = \frac{2}{5}\Delta_{HL}^2 \tau_2, \quad (25)$$

which, as expected, coincides with the non-model result of stochastic perturbation theory, Eq.(1). In general, one may conclude that in this regime the anisotropic part of instantaneous Luttinger Hamiltonian is self-averaged by rapid isotropic reorientations of the $M$-frame and the spherical symmetry of the system is restored; $J$ is a good quantum number and it is impossible to distinguish between the HH and the LH components of the $\Gamma_8$ quadruplet. We remark that in this limit the relaxation time of the $n$-th multipole moment in the hole density matrix can be readily obtained in analogy with a nuclear spin $I > 1$ systems[15]: $\tau_n^{-1}(\Gamma_8) = n(n+1)[14-n(n+1)]\Delta_{HL}^2 \tau_2 / 60$ ($n \leq 3$).

All of the above results are obtained under assumption that the reorientation of the $M$-frame is isotropic, $f(\widetilde{\Omega}) = f(\cos\widetilde{\beta})/(4\pi^2)$. Clearly, this model is not appropriate for charge carriers confined in low-dimensional nanostructures. Consider, for example, a quantum well or a QD grown in the [001] direction. In this case, there is a negligible mixing between the HH and LH subbands. The momentum states along this axis are quantized and characterized by the subband number $n_z$. Within the hard-wall approximation, this leads to the following redefinition of the parameter $D_k$ in Eq.(2): $\tilde{D}_k := -(2\pi^2 n_Z^2 / a^2 - k_{x_M}^2 - k_{y_M}^2)/2$, where $a$ is the height of the well. Consequently, for very thin nanostructures, $a < 5$ nm, the size quantization significantly amplifies the splitting between HH and LH subbands and a hole motion is constrained to the plane perpendicular to the growth direction. Within the framework of the theory presented here, this situation can be described by Eq.(19) with $f(\tilde{\Omega}) = f(\tilde{\alpha})\delta(\cos\tilde{\beta} - 1)\delta(\tilde{\gamma})$, where $f(\tilde{\alpha})$ determines the degree of correlation between successive in-plane angular jumps of the $M$-frame about $z_M$ that coincides with [001] axis. Taking into account that $f(\tilde{\alpha}) = f(-\tilde{\alpha})$, we obtain $W_{qq'}^l = \tau_Q^{-1}\delta_{qq'}[1-\delta_{q0}]$. As a result, the fundamental submatrix $<K,l,q=0|-i(\omega\hat{1}+\hat{\Lambda})+\hat{\Gamma}|K',l',q'=0>$, for any $K$ and $l$, contains only $W_{00}^l = 0$. Thus, if $\vec{J}$ is initially parallel to [001] axis of a quantum well or a QD, stochastic *in-plane* reorientations of the $M$-frame will not perturb this orientation.

### III. Conclusion

The EY mechanism is well established for electrons[11] and is related to the action of the spin-orbit-rotation interaction that emerges due to rotational perturbation of the Bloch wave-functions in the course of an *adiabatic* collision[25][26][27]. This process leads to spin-dephasing with characteristic time proportional to the orientational relaxation time of the electron crystal momentum. For holes, however, as long as $\Delta_{HL}^2 \tau_c^2 << 1$ the decay of spin polarization in bulk crystals is dominated by *nonadiabatic* intersubband HH - LH transitions, which strongly depend on the ratio between the $\Delta_{HL}$ and $\tau_2$. Hilton and Tang were able to distinguish between HH and LH bands. Thus, one may conclude that the system under study was in the slow modulation regime, $\Delta_{HL}^2 \tau_2^2 > 1$. In this limit, the relaxation kinetics is rather complex. It is characterized by dumped oscillations in the population of heavy and light holes described by the spectral function Eq.(21). The integral rate of the hole spin relaxation, Eq.(24), is associated with the orientational relaxation time of the effective quadrupole tensor and is not equivalent to the orientational relaxation time of the hole crystal momentum. The observed in Ref. [14] spin relaxation time of HHs was about 0.1 ps. Yet for GaAs, at room temperatures, the estimate based on Eq.(25) yields $\tau_J \approx 0.3$ ps, which is approximately three times longer than the experimental value. This discrepancy is not surprising, because the motional narrowing always leads to longer spin relaxation times. Our results suggest that unresolved HH and LH bands may be consistent with relatively long $\tau_J$. The absence of the hole optical orientation does not necessarily mean that the rates of angular and linear momentum relaxation are the same. Since $\Delta_{HL} \sim \gamma_2 k^2$, in the collision-dominated regime $1/\tau_J \sim \gamma_2^2 k^4$ and one may anticipate slower *J*-relaxation at lower temperatures and materials with smaller $\gamma_2$. Note, however, that similar to fluctuations of a crystal field splitting, the thermal fluctuations of the *magnitude* of an anisotropic part of the instantaneous Luttinger Hamiltonian will adiabatically modulate the gap between HH and LH components of the $\Gamma_8$ quadruplet. This process also leads to a pure spin dephasing[28]. To keep calculations and the results as simple as possible, in this paper we ignore the adiabatic dephasing. This approach allows focusing on the effect of nonadiabatic modulation of $\Delta_{HL}$, which provides a very efficient dumping mechanism of coherent HH - LH oscillations in bulk crystals.

Our results clearly demonstrate that the DP-like quadrupole mechanism of hole spin relaxation is suppressed by the 2D-confinement of the carrier's motion and increased subband splitting in low-dimensional semiconductor nanostructures. Note that due to Van Vleck cancellation, other SOC-related mechanisms of hole spin relaxation in low-dimensional nanostructures become also insignificant at zero magnetic fields[29]. Moreover, the splitting of the Berry connection for states with the helicity difference $\Delta m > 1$, makes the adiabatic EY mechanism ineffective for HHs in axially symmetric 2D [001] nanostructures[25][30]. Hence, at zero magnetic fields the nonadiabatic stochastic modulation of SOC-induced effective magnetic fields in systems with broken inversion symmetry[12][13] should dominate the effective spin-1/2 relaxation in widely separated HH and LH subbands in QDs and quantum wells.

**Appendix**.

Let us rewrite Eq.(7) in the following form

$$K_{J_0}(t) = Tr \int D(\Omega) \rho_{eq}^{(L)} J_0^{(L)}(\Omega,t) J_0^{(L)} D(-\Omega) \, d\Omega. \quad (A1)$$

Utilizing the transformation low

$$J_0^{(M)} = D(\Omega) J_0^{(L)} D(-\Omega) = \sum_q D_{q0}^1(\Omega) J_q^{(L)}, \quad (A2)$$

we can reduce Eq.(A1) to

$$K_{J_0}(t) = Tr \rho_{eq}^{(M)} \sum_q \widetilde{J}_q^{(M)}(t) J_q^{(L)}, \quad (A3)$$

where we introduce the designation, see Eq.(10),

$$\tilde{J}_q^{(M)}(t) = \int D(\Omega) J_0^{(L)}(\Omega,t) D(-\Omega) D_{q0}^1(\Omega) d\Omega \qquad (A4)$$

Note that operator $\tilde{J}_q^{(M)}(t)$ is determined in the *M*-frame. Now we shall see that the property of the kernel Eq.(8) is sufficient to derive the kinetic equation that is closed with respect to this operator. For that purpose, multiply the LHS of Eq.(5) by $D(\Omega)$ and its RHS by $D(-\Omega)D_{q0}^1(\Omega)$. Then integration over $\Omega$ yields

$$\dot{\tilde{J}}_q^{(M)}(t) = i\hat{L}_{k^2}^{(M)}(0)\tilde{J}_q^{(M)}(t) - \tau_Q^{-1}[\tilde{J}_q^{(M)}(t) - \\ - \int d\Omega \int d\Omega' f(\Omega-\Omega')D(\Omega)J_0^{(L)}(\Omega',t)D(-\Omega)D_{q0}^1(\Omega)] \qquad (A5)$$

Representing the integral term of Eq.(A5) as

$$\sum_{q_1}\int d\Omega \int d\Omega' f(\Omega-\Omega')D(\Omega-\Omega')[D(\Omega')J_0^{(L)}(\Omega',t)D(-\Omega')D_{q_10}^1(\Omega')]D(\Omega'-\Omega)D_{qq_1}^1(\Omega-\Omega')$$

and taking into account that integration over $\Omega$ is equivalent to integration over $\tilde{\Omega} = \Omega - \Omega'$ (Jacobian of this transformation equals 1), we obtain the following kinetic equation closed relative to the operator $\tilde{J}_q^{(M)}(t)$:

$$\dot{\tilde{J}}_q^{(M)}(t) = i\hat{L}_{k^2}^{(M)}(0)\tilde{J}_q^{(M)}(t) - \tau_Q^{-1}[\tilde{J}_q^{(M)}(t) - \sum_{q_1}\int d\tilde{\Omega} f(\tilde{\Omega})D(\tilde{\Omega})\tilde{J}_{q_1}^{(M)}(t)D(-\tilde{\Omega})D_{qq_1}^1(\tilde{\Omega})]$$

Taking into account that in isotropic 3D media $\varphi(\Omega) = 1/8\pi^2$, the initial condition to this equation can be easily derived from Eq.(6), (A2), and (A4): $\tilde{J}_q^{(M)}(t) = J_q^{(L)^+}/3$.

---

[1] T. Jungwirth, J. Sinova, J. Masek, J. Kucera, and A. H. MacDonald, Rev. Mod. Phys. **78**, 809 (2006).

[2] J. Wang, et al., Phys. Rev. Lett. **95**, 167401 (2005).

[3] D. Heiss et al., arXiv:0705.1466 (2007).

[4] J. Wunderlich, et al., Phys. Rev. Lett. **94**, 047204 (2005).

[5] V. Cerletti, W. A. Coish, O. Gywat, and D. Loss, Nanotechnology **16**, R27 (2005).

[6] W. A. Coish, V. N. Golovach, J. C. Egues, and D. Loss, Physica Status Solidi (b) **243**, 3658 (2006); D. Klauser, W. A. Coish, and D. Loss, Advances in Solid State Physics, v. 46 (2006).

[7] D. Culcer, C. Lechner, and R. Winkler, Phys. Rev. Lett. **97**, 106601 (2006).

[8] Yu. A. Serebrennikov, Phys. Rev. B. **71**, 233202 (2005).

[9] D. Culcer and R. Winkler, Phys. Rev. B. **76**, 195204 (2007).

[10] T. Uenoyama and L. J. Sham, Phys. Rev. Lett. **64**, 3070 (1990); R. Ferreira and G. Bastard, Europhys. Lett. 23, 439 (1993).

[11] I. Zutic, J. Fabian, and S. Das Sarma, Rev. Mod. Phys. **76**, 323 (2004).

[12] N. S. Averkiev, L. E. Golub, and M. Willander, J. Phys.: Condens. Mat. 14, R271 (2002); B. A. Glavin and K. W. Kim, Phys Rev. B **71**, 035321 (2005).

[13] C. Lu, J. L. Cheng, and M. W. Wu, Phys. Rev. B **73**, 125314 (2006).

[14] D. J. Hilton and C. L. Tang, Phys. Rev. Lett. **89**, 146601 (2002).

[15] A. Abragam and R. Pound, Phys. Rev. **92**, 943 (1953).

[16] J. D. Wiley, Phys. Rev. B **4**, 2485 (1971).

[17] A. I. Burshtein and Yu. S. Oseledchik, Sov. Phys. JETP **51**, 1072 (1966).

[18] R. Kubo, Advan. Chem. Phys. **16**, 101 (1969); J. Freed, in *Electron-Spin Relaxation in Liquids*, edited by L. Muus and P. Atkins (Plenum, New York, 1972); A. I. Burshtein, A. A. Zharikov, and S. I. Temkin, J. Phys. B **21**, 1907 (1988); D. Schneider and J. Freed, Adv. Chem. Phys. **73**, 387 (1989).

[19] J. M. Luttinger, Phys. Rev. **102**, 1030 (1956).

[20] N. O. Lipari and A. Baldereschi, Phys. Rev. Lett. **25**, 1660 (1970).

[21] R. Winkler, *Spin-Orbit Coupling Effects in Two-Dimensional Electron and Hole Systems* (Springer, Berlin, 2003)

[22] D. A. Varshalovich, A.N. Moskalev, and V.K. Khersonsky, *Quantum Theory of Angular Momentum* (World Scientific, Singapore, 1988).

[23] A. Thayer and A. Pines, Acc. Chem. Res. **20**, 47 (1987).

[24] Yu. A. Serebrennikov, Chem. Phys. Lett. **137**, 183 (1987); Yu. A. Serebrennikov and M. I. Majitov, Chem. Phys. Lett. **157**, 462 (1989); Yu. A. Serebrennikov, Adv. Magnetic and Optical Res., **17**, 47 (1992).

[25] Yu. A. Serebrennikov, Phys. Rev. Lett. **93**, 266601 (2004).

[26] Yu. A. Serebrennikov, Phys. Rev. B **73**, 195317 (2006).

[27] Yu. A. Serebrennikov, Phys Lett. A, DOI: 10.1016/j.physleta.2007.09.001; ArXiv: cond-mat/0708.2565.

[28] M. I. Dyakonov and A. V. Khaetskii, Sov. Phys. JETP **59**, 1072 (1984).

[29] D. V. Bulaev and D. Loss, Phys. Rev. Lett. **95**, 076805 (2005).

[30] Yu. A. Serebrennikov, ArXiv: cond-mat/0508.566.